\documentclass[twocolumn,prb,groupedaddress,showpacs,epsf]{revtex4}
\usepackage[dvips]{graphicx}

\begin{document}
\title{Quantum electromechanics: Quantum tunneling near resonance and
qubits from buckling nanobars}
\author{Sergey Savel'ev$^{(1,2)}$, Xuedong Hu$^{(1,3)}$, A. Kasumov$^{(4)}$, and Franco
Nori$^{(1,5)}$} \affiliation{$^{(1)}$ Frontier Research System,
The Institute of Physical and Chemical Research (RIKEN), Wako-shi,
Saitama, 351-0198, Japan \\
$^{(2)}$Department of Physics, Loughborough University,
Loughborough LE11
3TU, UK\\
$^{(3)}$ Department of Physics, University at Buffalo, SUNY,
Buffalo, NY 14260-1500, USA \\
$^{(4)}$ Laboratoire de Physique des Solides, Associ{\'e} au CNRS,
Universit{\'e} Paris-Sud, F-91405, Orsay, France\\
$^{(4)}$ Center for Theoretical Physics, Department of Physics,
University of Michigan, Ann Arbor, MI 48109-1040, USA}
\date{\today}
\begin{abstract}
Analyzing recent experimental results \cite{kasumov,rfsquid}, we
find similar behaviors and a deep analogy between three-junction
superconducting qubits and suspended carbon nanotubes. When these
different systems are ac-driven near their resonances, the
resonance single-peak, observed at weak driving, splits into two
sub-peaks (Fig.~1) when the driving increases. This unusual
behavior can be explained by considering quantum tunneling in a
double well potential for both systems. Inspired by these
experiments, we propose a mechanical qubit based on buckling
nanobars---a NEMS so small as to be quantum coherent.
To establish buckling nanobars as legitimate candidates for
qubits, we calculate the effective buckling potential that
produces the two-level system and identify the tunnel coupling
between the two local states.  We propose different designs of
nanomechanical qubits and describe how they can be manipulated.
Also, we outline possible decoherence channels and detection
schemes. A comparison between nanobars and well studied
superconducting qubits suggests several future experiments on
quantum electromechanics.
\end{abstract}
\maketitle

\section{Introduction}

Micro- and nano-electromechanical systems (MEMS and NEMS) have
attracted widespread attention because of their broad spectrum of
functionalities, their tiny sub-micron sizes, and their unique
position bridging microelectronic and mechanical
functions\cite{Cleland_book}.  Sophisticated tools ranging from
mirrors and sensors, to motors and multi-functional devices have
been fabricated \cite{Roukes,Munich}.  As the size of these
devices shrinks, experimental studies of NEMs are rapidly
approaching the quantum limit of mechanical oscillations
\cite{Munich,LaHaye,Schwab_th_quanta,Blencowe,Boston}, where
quantum coherence and superposition should result in quantum
parallelism and the possibility of information processing.

The emergence of quantum electromechanical devices (see, e.g.,
Ref.\cite{Cleland_book,Roukes,LaHaye,Schwab_th_quanta,Blencowe,Boston}
and references therein) brings both challenges, such as the
inevitable and ubiquitous quantum noise, and promises, such as
macroscopic quantum coherence \cite{Blencowe,Werner,Carr} or
quantum teleportation \cite{eisert,buks}. Indeed, during the past
several years the quantum mechanical properties of NEMS and how
they can be coupled to other quantum mechanical objects have been
very actively studied
\cite{Armour,Martin,Tian,nishiguchi,Kirschbaum,Bargatin}. The
experimental pursuit of these studies has so far focused on
cooling a device to reach states where quantum fluctuations in the
lowest energy eigenmode dominate over thermal fluctuations
\cite{Martin,LaHaye}.  Such eigenmodes are generally
harmonic-oscillator modes with equal energy-spacings and follow
bosonic statistics.

Among NEMS there also exist systems that can be well approximated
by two levels in the quantum limit, so that they might be
candidates for qubits.  For example, when a longitudinal strain
above a certain critical value is applied to a small bar with one
or two ends fixed, there exist two degenerate buckling modes, as
schematically shown in Fig.~2.  In the quantum mechanical limit,
these two modes represent the two lowest-energy states.  They can
be well separated from the higher-energy excited states, so that
at low temperatures the buckling nanobar can be properly described
as a two-level system.  Furthermore, since nanobars can be charged
or can carry electrical current, the electric or magnetic field
can be conveniently used to manipulate their quantum states.

The fascinating prospect of observing quantum coherent phenomena
in a macroscopic mechanical oscillator is a main motivation of
this study.  To achieve this goal, a NEMS needs to possess at
least two general attributes: small size and a high fundamental
frequency. A carbon nanotube is a natural candidate: it is very
thin (and can be very short) while still being stiff \cite{Treacy}
(thus having high vibration frequency) because of the strong
covalent bonds between carbon atoms within the graphene sheet
\cite{Wong,Poncharal}.
However, without strong experimental support, it might be hard to
judge the feasibility of achieving macroscopic quantum coherence
in NEMS. In contrast to NEMS, experimental evidence has
demonstrated that superconducting charge \cite{Nakamura}, flux
\cite{Nakamura1}, and phase \cite{phase-martinis} qubits, using
Superconducting Quantum Interference Device (SQUID), do exhibit
quantum coherence. Thus, it is important to look for possible
similarities in the behavior of superconducting qubits and NEMS.
This should help future prospects to observe quantum coherence in
NEMS.

\section{Splitting resonance as a manifestation of quantum
tunnelling in three-junction SQUIDs and NEMS}

Experiments \cite{kasumov} on suspended single-wall carbon
nanotubes (with diameter of about 20$\;$nm and length of about 1.7
$\mu$m) excited by an electromagnetic wave show a resonance peak
with an unusual shape (Fig.~1a), for one of the fundamental
harmonics $\omega_0$. As expected, a Lorentz-form resonance peak,
that grows with increasing intensity of the ac field, was observed
at weak drivings. Surprisingly, when further increasing the
amplitude of the externally applied electromagnetic wave, this
peak splits into two sub-peaks \cite{kasumov}. When further
increasing the ac drive, these two sub-peaks gradually move away
from each other, while their heights stop growing. It is important
to stress that this phenomenon was observed at a frequency
($\omega_0/2\pi\sim 2\;$GHz) and temperature ($T\sim 100\;$mK)
where quantum effects start to dominate over thermal noise
($T\lesssim \hbar\omega_0/k_B\approx 100$~mK, with Boltzmann
constant $k_B$). Also the dissipation in the system was quite low
(quality factor $Q\;\sim\;$1500), which is important to observe
quantum effects. Note that the fundamental frequency of this
device can be easily increased (at least by an order of magnitude,
well into the operating frequency 1-15$\;$GHz of many
superconducting qubits), for experimentally available carbon
nanotubes with shorter length.

Interestingly, a similar phenomenon has been recently
found\cite{rfsquid} for an Al three-junction SQUID qubit coupled
to a Nb resonant tank circuit  (Fig.~1b). It was experimentally
proven\cite{rfsquid}, via the observation of quantum hysteresis
(Landau-Zener transitions), that this circuit was operated in the
quantum regime (at $\omega_0/2\pi\sim 20\;$MHz and $T\gtrsim
10\;$mK), though for a worse ratio of quantum to thermal noise
compared to the carbon nanotubes \cite{kasumov} (i.e.,
$\omega_0/T\sim$ one order of magnitude higher for the nanotube).
When the magnetic flux in the SQUID was driven as $\Phi=\Phi_{\rm
dc}+\Phi_{\rm ac}\cos(\omega t)$, the resonance in the response,
probed via the tank voltage as a function of the dc flux, was
found \cite{rfsquid} to exhibit a transformation from a
Lorentz-form single-peak to a double-peak shape (Fig.~1b) in
striking similarity to suspended driven carbon nanotubes
\cite{kasumov}. Sweeping the dc flux in the SQUID corresponds to
changing the fundamental frequency as
\begin{equation}
\omega_0=\omega_0(\Phi_{\rm dc }=0)+\Delta\omega_0(\Phi_{\rm dc}),
\end{equation}
with $\Delta\omega_0\propto (\Phi_{\rm dc}-\Phi_0/2)$ and flux
quantum $\Phi_0$. Thus, we find that the
measured\cite{kasumov,rfsquid} response, near resonance, of both
systems (carbon nanotube and three-junction SQUID qubit) depends
on the difference $\omega-\omega_0$. This will be useful to
establish below that the measurements for driven carbon
nanotubes\cite{kasumov} and three-junction SQUID
qubit\cite{rfsquid} essentially probe the {\it same} effect.

\begin{center}
\includegraphics*[width=9cm]{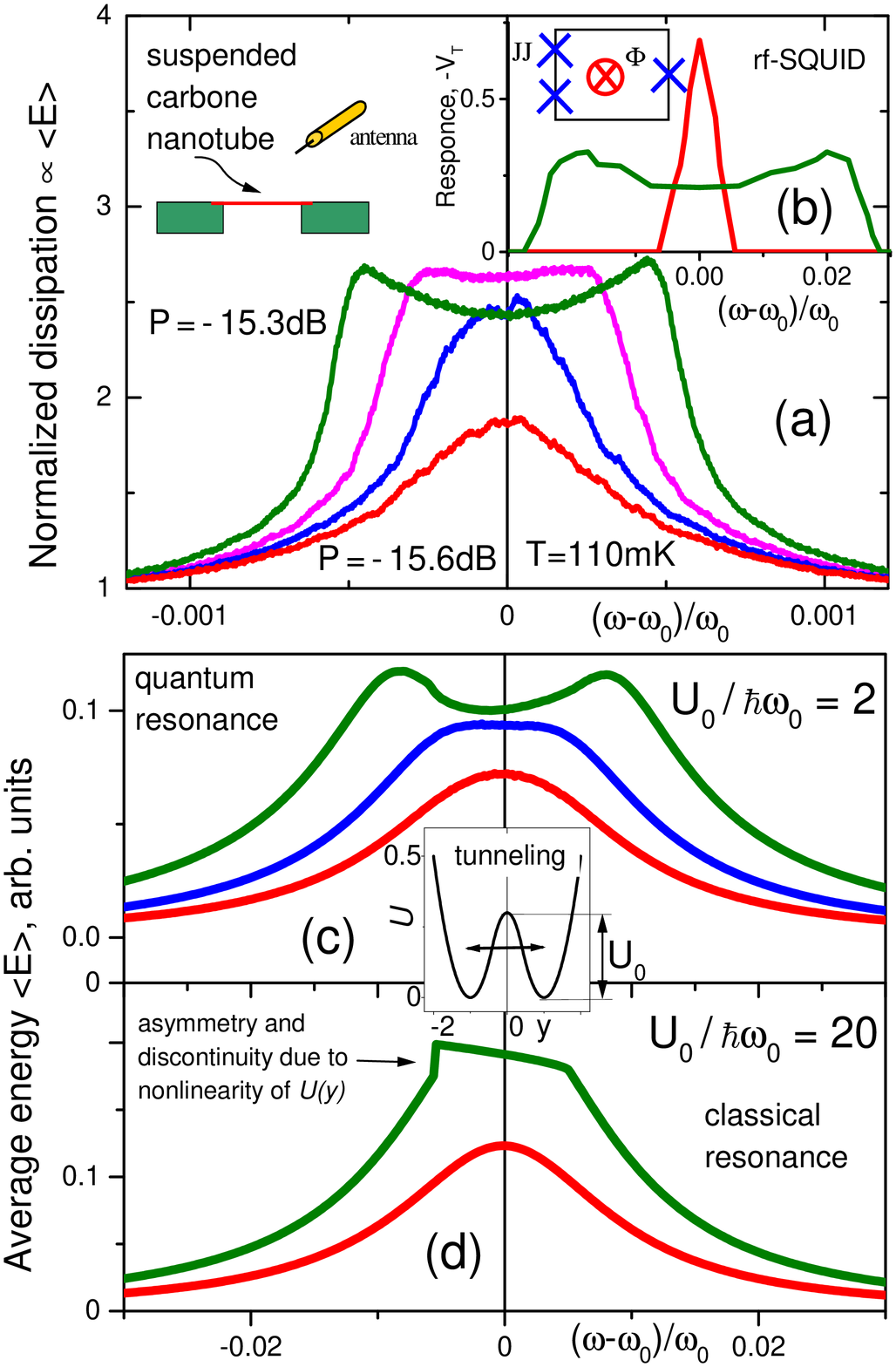}
\end{center}
\begin{figure}[!htp]

\vspace*{-0.5cm}
\caption{(a) Experimental resonance\cite{kasumov} of the effective
dissipation (which is proportional to the average energy $\langle
E\rangle$) as a function of the dimensionless frequency
$(\omega-\omega_0)/\omega_0$, for a suspended carbon nanotube, for
different intensities $P$ of the externally applied ac
electromagnetic wave. (b): the resonance in the response of the
three-junction SQUID probed by measuring\cite{rfsquid} the voltage
for the resonant tank $V_T$ for different driving amplitudes as a
function of $(\omega-\omega_0)/\omega_0$. For these different
systems, a similar splitting of the resonance peak was
observed\cite{kasumov,rfsquid}. Schematic diagrams for a driven
suspended carbon nanotube and three junction SQUID are also shown
at the top. The simulated resonance [average energy $\langle
E\rangle$ versus the dimensionless frequency
$(\omega-\omega_0)/\omega_0$] for quantum (c) and classical (d)
particles moving in the double-well potential shown in the inset
of (c). A quantum description agrees with
experiments\cite{kasumov,rfsquid}(Fig.~1a,b): For quantum
particles, the standard Lorentz-form resonance peak for weak ac
drives (c, red curve, $A=0.005$) splits into two sub-peaks due to
tunnelling at stronger drives (blue curve, $A=0.01$; green curve,
$A=0.0135$). A classical description cannot describe
experiments\cite{kasumov,rfsquid}(Fig.~1a,b): For classical
particles, the weak-driving Lorentz-form peak (red curve,
$A=0.005$) becomes asymmetric and exhibit discontinuous jumps
$E(\omega)$ (green curve, $A=0.0135$) due to the
nonlinearity of $U(y)$.} %\vspace*{-1cm}
\end{figure}
\newpage

We propose how the incorporation of quantum tunnelling can be used
to understand these still-unexplained experimental observations in
nanotubes\cite{kasumov} and three junction SQUIDs\cite{rfsquid}.
In order to interpret the splitting of the resonance peak (Fig. 1a
and 1b), we employ a quasi-classical model\cite{quasi} described
by the equation of motion
\begin{equation}
\ddot{y}\,+\,2\lambda\dot{y}\,+\,\partial U(y)/\partial y\;=\;
A\,\cos(\omega t),
\end{equation}
for the double-well potential shown in the inset in Fig.~1c, with
small damping $\lambda=0.01$, and different driving amplitudes
$A$. We assume that the particle (which mimics the buckling mode
or current in the SQUID) can tunnel from one well to the other
with transmission coefficient
\begin{equation}
D=\exp\left[-\frac{2\sqrt{2}}{\hbar}\int_{-a}^{a}\sqrt{U-E}\;dy\right],
\end{equation}
 where $a$
is the classical turning point, and $E=(\dot{y})^2/2+U(y)$ is the
energy of the particle. This minimal model allows to qualitatively
describe the resonance of a quantum particle in a double well
potential (a more complete, fully quantum mechanical, theory will
be published elsewhere \cite{rakh}) and also to obtain the
transition to a classical description (Fig.~1c,d).

When the potential barrier $U_0$ is comparable with
$\hbar\omega_0$, we find that the single resonance peak, at low
driving, splits into two sub-peaks (Fig.~1c) for higher drives; in
agreement with experimental findings \cite{kasumov,rfsquid}. The
physical origin of this effect is the following: (i) At low
driving the energy $E$ of the quantum particle (whether the SQUID
or the nanotube) is also low and the probability of tunnelling is
negligibly small, thus, the usual Lorentz-form of the resonance
occurs; (ii) When the driving (and, thus, energy) increases, the
particle starts to tunnel between the wells. A tunnelling event is
equivalent to an inversion around the origin, i.e., changing $y$
to $-y$: like replacing the left well by the right one, or vice
versa. As a result, right after a tunnelling event, the average
power ${\cal P}=\langle \dot{y} F\cos\omega t\rangle$ the particle
receives from the external force (e.g., the power absorption by
the nanotube from the external electromagnetic wave) changes its
sign ${\cal P}\;\rightarrow\;-\;{\cal P}$ and, hence, the energy,
which increases before tunnelling, starts to decrease temporarily.
Thus, the average energy $\langle E\rangle$ also decreases when
the number of tunnelling events increases. This explains the
minimum of $\langle E(\omega)\rangle$ at $\omega=\omega_0$
(instead of the maximum observed at weak driving) and also the
splitting of the resonance peak --- because the tunnelling
frequency sharply increases at the resonance frequency $\omega_0$.

For the case of ``more classical'' particles, $U_0\gg
\hbar\omega_0$, the probability of tunnelling is always very low.
When driving increases, the particle (which is always located in
the same well) begins to feel the strong nonlinearity of the
potential $U(y)$, resulting in the energy dependence of the
oscillation frequency $\omega_0(E)$. Instead of a split resonance,
the resonance peak as a function of frequency shows the standard
asymmetric shape with a sharp jump of $\langle E(\omega)\rangle$,
associated with mechanical hysteresis\cite{mechanics} (Fig.~1d).

Therefore, we conclude that the two-peak resonance indicates that
both the three-junction SQUID qubit\cite{rfsquid} and the
suspended carbon nanotube \cite{kasumov}, operate in the quantum
regime. Thus, the technology for fabricating suspended buckled
carbon nanotubes, working as nanomechanical qubits, already
exists\cite{kasumov}.  In view of the explosive growth of NEMS
technology, below we discuss the prospect of such buckling charged
nanobars (the clamping at the base ensures an anisotropic nanobar
instead of an isotropic nanotube) as candidates of quantum bits
for quantum information processing. Indeed, previous work on
quantum buckling (e.g., \cite{Werner,Carr}) were not focused on
operating these as qubits, which is our focus here \cite{sci-am}.

\section{Controlling quantum states of a nanorod in a double-well
potential.}

The double-well potential (Fig.~2a) corresponding to the buckling
modes $y\sin[\pi l/l_{\max}]$ of a nanorod (Fig.~2b) takes the
form:
\begin{equation}
U(y)=\alpha y^2+\beta y^4+\frac{2l_{\max} f_\perp}{\pi}y,
\end{equation}
where the parameters can be estimated \cite{ourprb} as
\begin{eqnarray}
\alpha(f)  &=& \frac{\pi^2}{4l_{\max}}(f_c-f), \nonumber \\
\beta(f) &=& \frac{\pi^4}{64l^{3}_{\max}} \left(
4f_c%\frac{4IE\pi^2}{l^{2}_{\max}}
- 3f \right),\nonumber \\ f_c  &=& \frac{I\;Y\pi^2}{l_{\max}^2}.
\label{potential}
\end{eqnarray}
Here, we introduce the tube length $l$ ($0\leq l\leq l_{\max}$),
the elastic modulus $Y$ and the moment of inertia $I$. To control
the rod we need: (i) a longitudinal compressing force $f$ acting
on the rod ends, and (ii) a transverse force $f_\perp$, which can
be produced via, e.g., interacting the charged nanorod with an
electric field.

For zero transverse force $f_\perp=0$, our qubit (Fig.~2b) is in
its degeneracy point. Indeed, this potential has two minima, at
$y\;=\;\pm\; y_0(f)=\pm\sqrt{\alpha/2\beta}$ that are separated by
a potential barrier
\begin{equation}\Delta U(f)\; =\; \alpha^2/4\beta.\end{equation}
The first two energy levels $E_1$ and $E_2$ in the right well can
be estimated assuming a parabolic potential well shape $U\approx
m\omega_0^2 (y-y_0)^2/2$ with
\begin{equation}
\omega_0(f) = \left(\frac{U''(y_0)}{m}\right)^{1/2} =
2\left(\frac{\alpha(f)}{m}\right)^{1/2}
\end{equation}
 and the mass $m$ of the
nanorod. Thus, we obtain
\begin{equation}
E_1(f) = 3\hbar(\alpha(f)/m)^{1/2}
\end{equation}
 and
\begin{equation}
 E_2(f)=5E_1(f)/3.
 \end{equation}
Due to quantum tunnelling between the left and right potential
wells, $E_1$ splits into two levels
\begin{equation}
E_1^{\pm}=E_1\pm\hbar\Delta_t.
\end{equation}
The tunnelling rate $\Delta_t$ between the left and right buckled
states is: \cite{landau-quantum}
\begin{equation}
\Delta_t(f) \;\approx\; \frac{2}{\pi} \sqrt{\frac{\alpha(f)}{m}}
\exp{ \left(-\;\frac{\pi \sqrt{2m} (\Delta U(f)-E_1(f))}{2 \hbar
\sqrt{\alpha(f)}}\right)}.
\end{equation}
It is important to stress that the longitudinal force $f$ allows
to control the tunnelling $\Delta_t(f)$ as well as the energy
levels itself $E_1(f)$. The higher levels are well separated from
the two lowest ones: $(E^+_1-E^-_1)/(E_2-E_1)\sim
\Delta_t/\hbar\omega_0 \ll 1$. Changing the transverse force
$f_\perp$ moves the system out of the degeneracy point, allowing
to manipulate the proposed nano-mechanical qubit.

\begin{figure}[!htp]
\begin{center}
\includegraphics*[width=7.7cm]{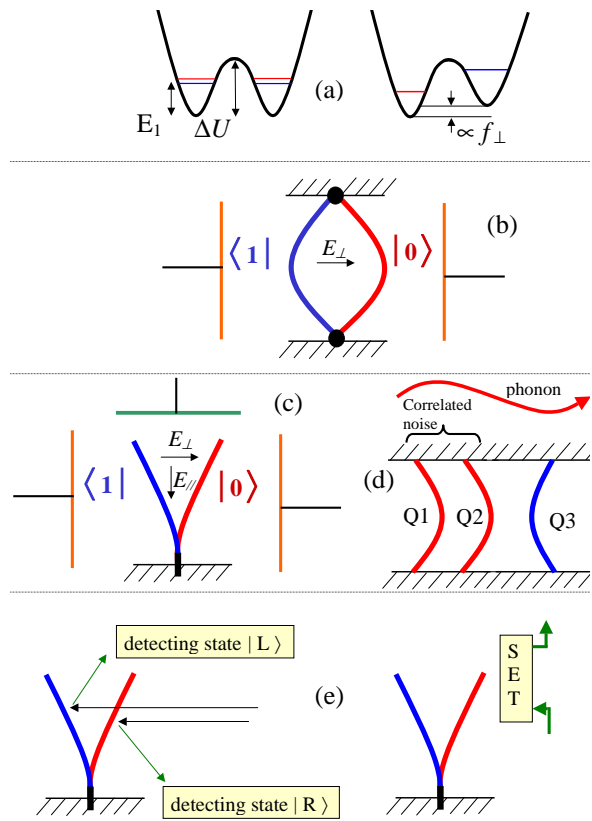}
\vspace*{-0.5cm}
\caption{(a) Double-well potential for a buckled nanobar. Due to
tunnelling from the right potential well to the left one, the
lowest energy level is split into two levels for $f_{\perp}=0$ as
shown in the left panel.  The lowest (blue) and the first excited
(red) levels correspond to the symmetric and antisymmetric
combinations of the wave functions localized in the left and right
potential wells.  The energy level splitting between the left and
right states can be controlled by the transverse force $f_\perp$
as shown in the top right panel. (b) A buckled rod qubit where the
compressed force applied to the rod ends controls the potential
shape [$\alpha$ and $\beta$ in Eq.~(\ref{potential})] and,
therefore, the energy splitting at the degeneracy point. The
transverse force $f_{\perp}$ allows to drive the bar to a
degeneracy point. (c) Another proposed design for a buckled-rod
qubit. In this case the possible decoherence originating from the
relative vibration of the top and bottom rod holders can be
avoided. By changing the charge of the top capacitor plate and the
charge of the rod itself it is possible to independently control
both the parallel-to-rod electric field $E_{\parallel}$ and its
gradient. This design allows a high level of control of both the
tunnelling and the energy splitting. (d) Correlated noise produced
by some phonons can be eliminated using a decoherence-free
subspace: qubits Q1 and Q2 can be associated with one logic qubit.
(e) Single-shot measurements could be done via either single
electron transistor (SET) or photon reflection.} %\vspace*{-1cm}
\end{center}
\end{figure}

\begin{figure}[!htp]
\begin{center}
\includegraphics*[width=9cm]{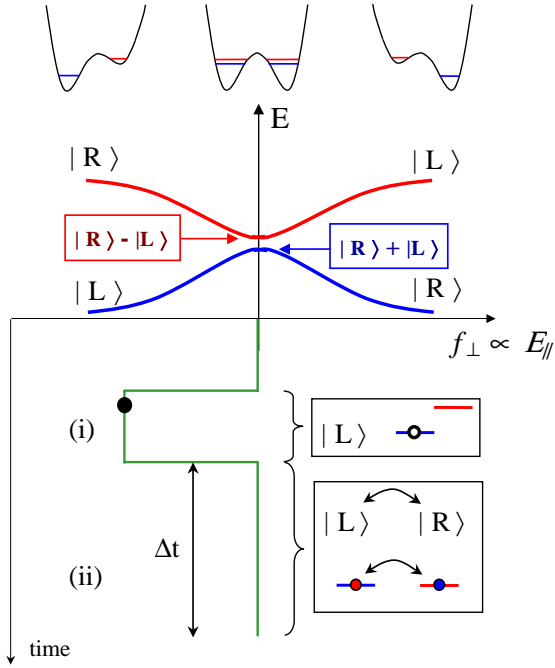}
\vspace*{-3.5cm}
\caption{A schematic diagram for controlling the quantum state of
a nanobar via coherent oscillations at the degeneracy point.
Tuning the system to its degeneracy point can be done by changing
the perpendicular electric field $E_{\perp}$. For the case when
$E_\perp$ and $E_\parallel$ is zero at $t=0$, the perpendicular
electrical field $E_\perp$ has to be applied first and then the
longitudinal field $E_\parallel$ (e.g., at the moment marked by
the solid circle on the $E_\perp({\rm time})$ curve). This
initializes the system during the stage (i). Switching off the
perpendicular electric field $E_{\perp}$ brings the system back to
its degeneracy point, and the nanorod starts to oscillate a time
$\Delta t$ during stage (ii). To observe coherent oscillations,
measurements (e.g., optically or electrically) must be done for
many values of $\Delta t$.}%\vspace*{-1cm}
\end{center}
\end{figure}

We now introduce another, more controllable, design of a nanorod
qubit, which is also less affected by decoherence coming from the
relative vibrations of the rod ends (Fig. 2c). In such a design we
consider a ``crossed'' electric field having a component
$E_{\parallel}$ along the rod and another $E_{\perp}$
perpendicular to the rod, which is clamped to a substrate only by
one end. Changing the charge of the top capacitor plate and the
charge of the rod itself can control both the electric field,
$E_{\parallel}$, on the rod and its gradient, $\partial
E_{\parallel}/\partial x$. By adjusting the different knobs, we
can tune the relative strength between thermal activation and
quantum tunneling, allowing the observation of transition between
these two regimes.

\section{Manipulation, decoherence, and detection of mechanical
qubits}

The manipulation of the mechanical qubits can be achieved
electrically (Fig. 3). For example, in analogy to the Cooper pair
box \cite{Nakamura} (see table), one can prepare the nanobar qubit
in the $|L\rangle$ state by setting a transverse electric field
towards the right (assuming the nanobar is negatively charged). By
{\it suddenly} turning off this electric field and bringing the
system to the degeneracy point, the nanobar state is prepared in a
coherent superposition of $(|R\rangle \pm |L\rangle)/\sqrt{2}$.
Because of the tunnel splitting, the system then starts to
oscillate coherently, with a frequency given by $\Delta_t$.
Therefore, by detecting the nanobar position, as a function of
$\Delta t$ (see Fig.~3), one can determine the coherent
oscillation frequency and the system decoherence. Driven coherent
transitions between the two qubit states can also be similarly
achieved.  A sinusoidal component can be added to the vertical
electric field that is used to control the tunnel coupling between
the $|L\rangle$ and $|R\rangle$ states, in analogy to the
microwave driving
force on the Josephson coupling in a Cooper pair box. %\cite{Vion}.
The study of free and driven coherent oscillations of a nanobar would help
clarify its quantum coherence properties and demonstrate its feasibility as a
qubit.

Universal quantum computing requires two-qubit operations.  For charged
nanobars, the inter-qubit interaction comes naturally in terms of the
electric dipole interaction between the bars, quite similar to the dipole
interaction used in other proposed qubits.  In the case of nanobar qubits,
one can again use the transverse electric field $E_T$ to tune selected qubits
into resonance, then apply microwaves to perform conditional rotations and
other operations.

\begin{figure}
\begin{center}
\vspace*{-1.5cm}
\hspace*{-0.5cm}\includegraphics*[width=9.5cm]{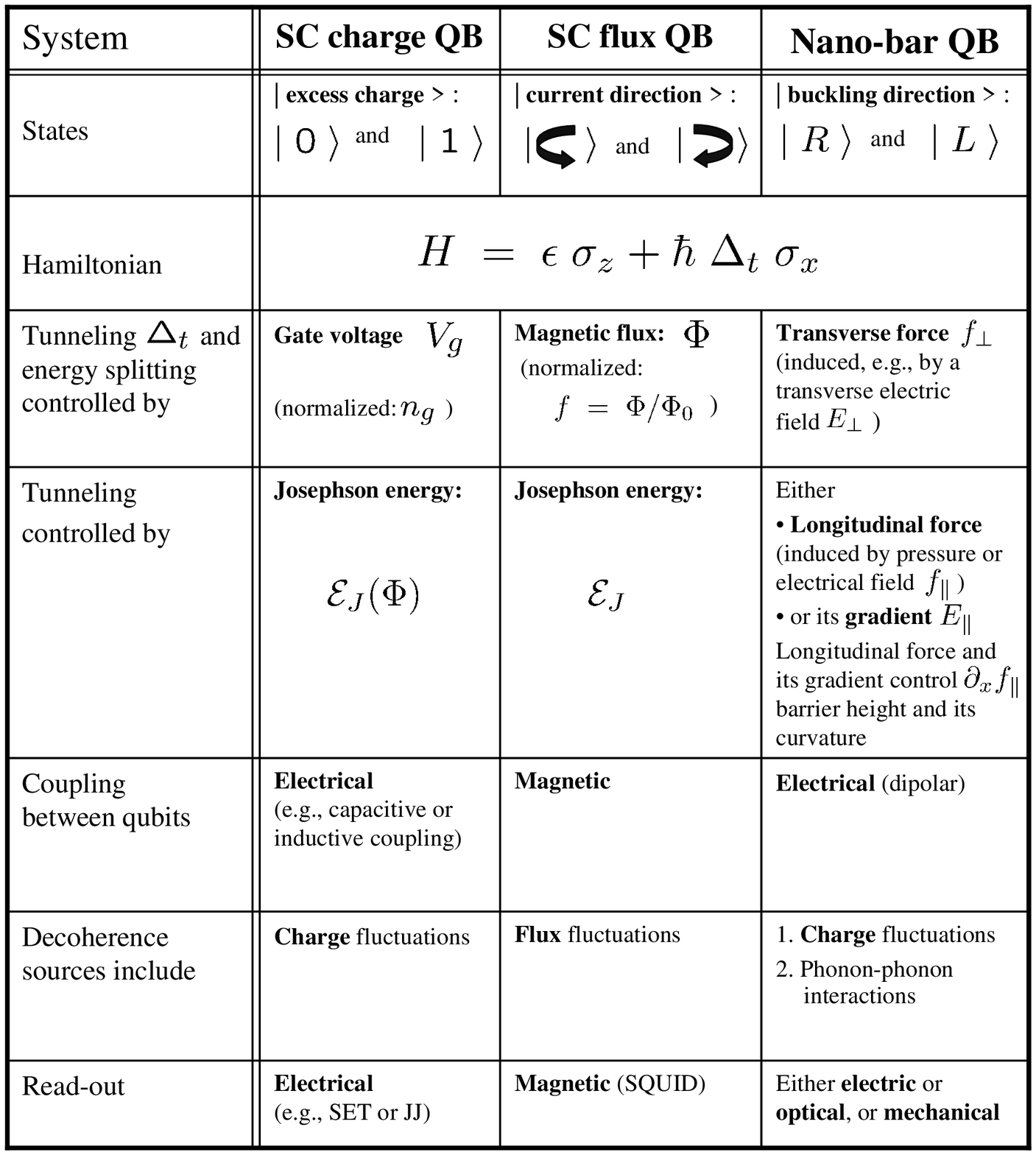}
\vspace*{-2cm}
\end{center}
%\vspace*{-2cm}
 Table: comparison of Josephson-junction
superconducting (JJ SC) charge, JJ SC flux and nano-bar qubits
(QB). \vspace*{-0.7cm}
\end{figure}

For NEMS, and for more conventional applications such as
resonators, major sources of noise \cite{Nowick,Cleland} (that
limit the quality factor $Q$) are internal thermomechanical noise
(such as heat flows due to inhomogeneous distribution of strain),
Nyquist-Johnson noise from the driving electrical circuit,
adsorption-desorption noise when a resonantor is moving in a
non-vacuum, noise from moving defects, etc. In the quantum
mechanical limit, some of these noises become unimportant.  For
example, there should be no heat flow: the nanobar should not be
driven beyond its first excited state. The NEMS can be placed in a
vacuum to reduce adsorption-desorption noise. The whole system,
including its electrical components, has to be cooled in a
dilution refrigerator in order to reach the two lowest states of
the nanobar, so that the Nyquist-Johnson noise is also suppressed.
The main source of quantum mechanical decoherence might be
internal dissipation caused by phonon-phonon interactions.  Since
the nanobars are clamped onto a much larger substrate, in which
lower energy phonon modes exist, coupling at the base of the
nanobars could lead to relaxation/excitation of the qubit states.
However, the long-wave-length phonons produce a correlated noise,
which can be reduced by working within a properly chosen
decoherence-free subspace (see schematics in Fig. 2d and, e.g.,
\cite{rakh,dfs}).
Another source of decoherence would be charge fluctuations.
Charged nanobars allow increased maneuverability of the NEMS; but
the surrounding and internal charge fluctuators can produce charge
noises that affect the relaxation and dephasing of the qubit.

Single-wall nanotubes are a natural candidate for the mechanical
qubit we consider here.  However, our proposal is not limited to
them. Indeed, a multiple-wall nanotube could provide a
higher-frequency mechanical oscillator and more favorable
condition for observing macroscopic quantum tunnelling and the
coherent evolution of mechanical motion. %, assuming that the motion
%between walls can be effectively pinned or reduced.
Silicon-based systems provide another enticing alternative,
especially from the perspective of fabrication. While it might be
difficult to clamp several single-wall nanotubes to make identical
nanobars (with controlled inter-bar distance, same length, same
buckling orientation, and same response to external stress),
fabricating a lithographically-patterned silicon-based
nanostructure would be much more reliable.  SiC nanobars have
shown higher stiffness \cite{Wong} compared to Si.  Unless
nano-assembled nanotubes can be made with sufficiently high
precision, materials that can be lithographically fabricated seem
more promising candidates for a larger-scale mechanical quantum
information processor.  This would be an ironic turn of events,
given that the first computers (by C. Babbage) were mechanical.

The detection of the nanobar state can be done either electrically
\cite{Knobel,LaHaye,Kirschbaum} or optically \cite{Mamin,kiyat}.
It has been shown in recent experiments \cite{LaHaye} that single
electron transistors are very sensitive to small charge
displacements.  Optical detectors can also be used, where light
scattering can detect the state of the bent nanorod (Fig. 2e).

Very recently, Ref.~\cite{badzey} fabricated suspended nanobars
(driven by a 25 MHz current through an attached electrode)
switching between two distinct states.  These suspended nanobars
have already been tested \cite{badzey} as very fast and very
low-power-consumption storage memory devices. Still, many
challenges lie ahead on the road to practical quantum
electromechanics.  We hope that our proposal here stimulate more
research towards the ultimate quantum limit of NEMS.

\section*{ACKNOWLEDGMENT}

This work was supported in part by the National Security Agency
(NSA) and Advanced Research and Development Activity (ARDA) under
Air Force Office of Research (AFOSR) contract number
F49620-02-1-0334; and also supported by the US National Science
Foundation grant No.~EIA-0130383.

\nobreak

\end{document}